# High-performance chiral all-optical logic gate based on topological edge states of valley photonic crystal

Xiaorong Wang ,[1,2] Hongming Fei,[1,2,*] Han Lin,[3,*] Min Wu,[1,2] Lijuan Kang,[1,2] Mingda Zhang,[1,2] Xin Liu,[1,2] Yibiao Yang,[1,2] and Liantuan Xiao[1,2]

[1]*College of Physics, Taiyuan University of Technology, Taiyuan 030024, China*
[2]*Key Laboratory of Advanced Transducers and Intelligent Control System, Ministry of Education, Taiyuan University of Technology, Taiyuan 030024, China*
[3] *School of Science, RMIT University, Melbourne, Victoria 3000, Australia and The Australian Research Council (ARC) Industrial Transformation Training, Centre in Surface Engineering for Advanced Materials (SEAM), Swinburne University of Technology, Hawthorn, Victoria 3122, Australia*
*\*feihongming@tyut.edu.cn; han.lin2@rmit.edu.au.*

**Abstract**: For all-optical communication and information processing, it is necessary to develop all-optical logic gates based on photonic structures that can directly perform logic operations. All-optical logic gates have been demonstrated based on conventional waveguides and interferometry, as well as photonic crystal structures. Nonetheless, any defects in those structures will introduce high scattering loss, which compromises the fidelity and contrast ratio of the information process. Based on the spin-valley locking effect that can achieve defect-immune unidirectional transmission of topological edge states in valley photonic crystals (VPCs), we propose a high-performance all-optical logic OR gate based on a VPC structure. By tuning the working bandwidth of the two input channels, we prevent interference between the two channels to achieve a stable and high-fidelity output. The transmittance of both channels is higher than 0.8, and a high contrast ratio of 28.8 dB is achieved. Moreover, the chirality of the logic gate originated from the spin-valley locking effect allows using different circularly polarized light as inputs, representing "1" or "0", which is highly desired in quantum computing. The device's footprint is $18 \times 12$ μm$^2$, allowing high-density on-chip integration. In addition, this design can be experimentally fabricated using current nanofabrication techniques and will have potential applications in optical communication, information processing, and quantum computing.

## 1. Introduction

Due to the development of augmented reality (AR)/virtual reality (VR) and big data, the increase in information traffic demands large bandwidth and reliable information processing. Optical communication and information processing have the advantages of large bandwidth, fast speed, high energy efficiency, and low heat generation. An all-optical network [1] is that the signal always exists in the form of light during the propagation and information processing, which overcomes the "electronic bottleneck" phenomenon caused by the previous optical-electrical-optical conversion [2]. It is necessary to develop an entire family of different all-optical components for all-optical networks, such as waveguides, wavelength division multiplexing (WDM) devices, optical delay lines, logic gates, and so forth. Currently, information processing is realized by electrical logic gates (AOLGs) based on electro-optic conversion all the time[3,4] due to the limited performance of current AOLG designs that cannot meet the requirements in applications. As a result, the speed is ultimately limited by the relatively slow electric logic gates. Therefore, designing high-performance AOLGs has become a stringent request and is key to realizing all-optical information processing.

An ideal logic gate performs operations on Boolean values, which should dissipate no power and change states instantaneously, similar to a step function. Such gates operate on discrete logic inputs and should have no propagation delay. For AOLGs, the propagation delay is neglectable due to the high speed of light. In addition, power dissipation can be minimized by achieving high transmittance. Furthermore, for photonic integrated circuit (PIC) operation, AOLGs should have a miniaturized footprint for high-density integration. In general, AOLGs that work with different spin states of photons are required for quantum computing. Therefore, AOLGs should have chiral responses to work with spin-up (right-handed circularly polarized, RCP) and spin-down (left-handed circularly polarized, LCP) photons.

Many designs of AOLGs based on different working principles have been demonstrated, such as semiconductor optical amplifier (SOA) [5-7], periodically poled lithium niobate (PPLN) waveguide [8,9], ring resonators [10-12], and photonic crystals (PCs) [13-15]. The performance of SOA-based AOLGs is limited by spontaneous emission noise and high integration complexity. In comparison, the AOLGs based on semiconductor ring resonators have the advantages of simplicity and low input power. But the speed is slow. Meanwhile, the AOLGs based on PPLN waveguides have fast switching speeds and low spontaneous emission noise. However, those AOLGs strongly depend on the working temperature and the incident light's polarization states, which hinder their applications. All the AOLGs mentioned above have relatively large footprints, which makes them unsuitable for high-density on-chip integration. In comparison, the footprint of AOLGs can be significantly reduced by using nanostructures, in which PC structures [16-25] are the most widely used. Different effects and principles of PC structures have been applied to design AOLGs, such as self-collimation [26,27], multimode interference [28,29], interference waveguides [30,31], and nonlinear effects [32,33]. However, all the designs above have relatively low transmittance due to the scattering loss introduced by structural defects. In addition, none of the AOLG designs, including all conventional and PC designs, can work with circularly polarized light (CPL) to maintain the spin states, which is important for quantum computing [41,42].

The recent development of topological photonic crystals (TPCs) [34-38] allows defect-immune unidirectional transmission of topological edge states, which opens new possibilities in designing photonic devices. Among different TPC designs, valley photonic crystals (VPCs) can be achieved based on dielectric materials without an external magnetic field, thus pushing the working wavelength to the telecommunication and the visible region [39,40]. Here based on the unique spin-valley locking effect of VPCs, we design a high-performance integratable all-optical logic OR gate based on the tunable topological edge states of VPCs. We tune the working bands of the two input channels to minimize their interference, achieving stable and high-fidelity outputs. The spin-valley locking effect allows robust unidirectional transmission of CPL. The transmittance of both channels is higher than 0.8, which meets the requirements of an ideal all-optical logic gate. In addition, an ultrahigh contrast ratio of 28.8 dB is achieved. Currently, the contrast ratio of AOLGs based on conventional PC structures is mainly in the range of 10-15 dB. So our design has a much higher contrast ratio than those AOLGs in recent references[30]. Furthermore, the chiral responses of the AOLGs allow using CPL with different handedness as logic inputs (logic "1" or logic "0"), which meets the requirements of quantum computing [40,41]. The footprint of the entire AOLG device is as small as $18 \times 12$ $\mu m^2$, allowing high-density on-chip integration. The device can be manufactured by the mature complementary-metal-oxide-semiconductor (CMOS) nanofabrication technique and will find broad applications in optical information processing and quantum computing.

## 2. Design of the all-optical OR gate

An OR logic gate (symbolized by "||" or "+") outputs a logic "1" if one or both the inputs are logic "1"; otherwise, it outputs a logic "0". Here, the logic "1" in an AOLG is defined as the input or the output having an incident or transmitted optical wave with a transmittance higher than 0.7. And the logic "0" is determined when the criterion is not met. According to this definition, an all-optical OR gate based on a VPC structure is designed and shown in Fig. 1(d). The gate comprises two mirror-symmetrical VPC structures, consisting of two input waveguides (INA and INB) and two output waveguides (OUT1 and OUT2). To minimize the interference between the two inputs, we tune the working bands of INA (centered at $\lambda_1$, Figs. 1(a) and (c)) and INB (centered at $\lambda_2$, Figs.1(b) and(c)) by adjusting the radii of the lattice at the boundary (highlighted by different colors in Fig. 1(d)). Meanwhile, the working band of the output waveguide OUT1 can cover the two working bands of INA and INB. In contrast, the OUT2 is designed to have extremely low transmittance in those working bands to prevent the undesired loss through the waveguide to maximize the transmittance through OUT. The schematics in Fig.1 show the four possible cases of the AOLG, namely 1 || 0 = 1, 0 || 1 = 1, 1 || 1 = 1, and 0 || 0 = 0. The two numbers on the left side correspond to the inputs from INA and INB, respectively. The number on the right side shows the output of OUT1.

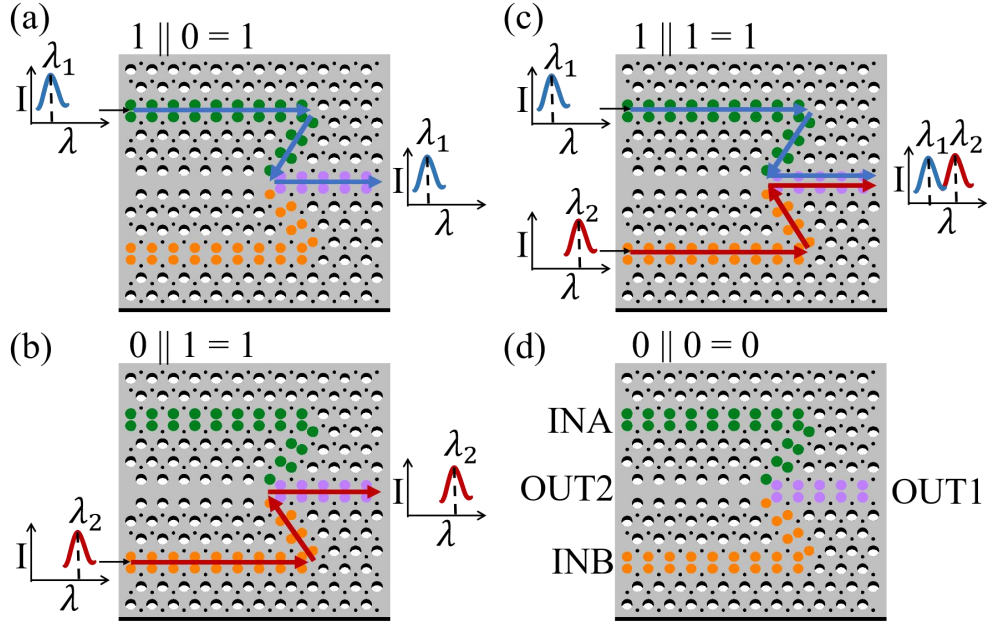

**Fig. 1.** The working diagrams of an all-optical OR logic gate : (a) Light inputs only from INA, as 1 || 0 = 1; (b) Light inputs only from INB, as 0 || 1 = 1; (c) Light inputs from both INA and INB, as 1 || 1 = 1; (d) Light inputs from neither INA nor INB, as 0 || 0 = 0.

The design process started with creating a honeycomb PC structure (Fig. 2(a)) composed of circular air holes embedded in a free-standing silicon substrate. The thickness of the free-standing silicon substrate (the dispersive refractive index of pure silicon is used) is $h = 220$ nm. The unit cell of this structure is composed of two sets of air holes (A and B) with the same radius ($r_A = r_B = 80$ nm), which has a $C_{6V}$ rotational symmetry. The lattice constant is $a = 450$ nm. Here we use a commercial three-dimensional (3D) time-domain finite difference (FDTD) software (Lumerical FDTD Solutions) to calculate the photonic band structure. The band diagram of the transverse electric (TE) mode is shown in Fig. 2(c) by the dashed lines, which show a Dirac point. Then the $C_{6V}$ symmetry is reduced to $C_{3V}$ symmetry by simultaneously increasing $r_A$ (to 120 nm) and decreasing $r_B$ (to 40 nm) to introduce a topological photonic bandgap (1410 nm - 1676 nm) due to the degeneracy of the K and K' valleys, indicated by the blue dot lines in Fig. 2 (c). In this way, VPC1 is designed (left panel in Fig. 2(b)), which can be converted to VPC2 (right panel in Fig. 2(b)) by applying mirror symmetry operation. The values of the Berry curvature and topologically invariant valley Chern number $C_V$ corresponding to the two VPCs were calculated, which are $C_V = -1$ for VPC1 and $C_V = 1$ for VPC2 [39], respectively.

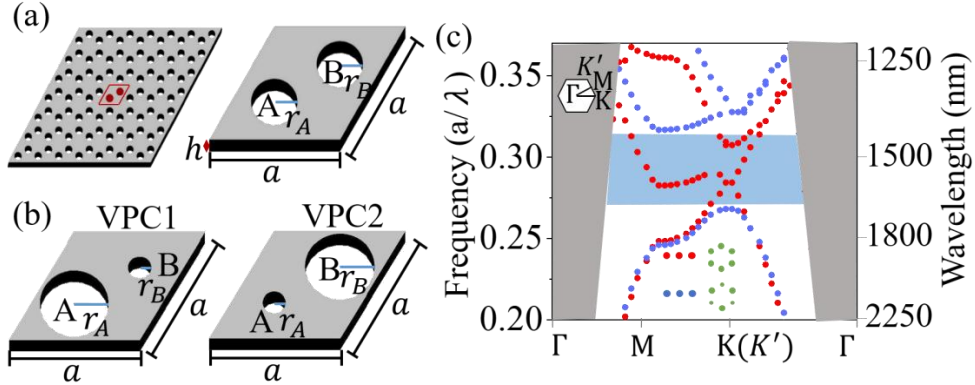

**Fig. 2.** Structure parameters and photonic band diagrams of PCs. (a) 3D schematics of the honeycomb PC structure; (b) 3D schematics of VPC1 and VPC2; (c) The photonic band diagram of the honeycomb PC (red dot line) and VPC1(blue dot line), the blue shaded area marks the bandgap; the gray region marks the air cone.

## 3. Tuning of topological edge states

There are two possible types of boundaries, the zigzag-shape and beard-shape boundaries, which can support topological edge states. The study of the edge states of the two boundaries shows that the working bandwidth of the beard-shape boundary is narrower than the bandgap, which allows tuning the position of the working bands within the bandgap. (The detail is shown in Supplementary material S1). In this study, we choose the beard-shape boundary as required to achieve different working bands within the bandgap. Depending on the small or large circles at the boundary, two kinds of beard-shape boundaries can be constructed: S-interface and L-interface. Due to the high transmittance of the L-interface, it is chosen for INA, INB, and OUT1. Fig. 3(a) shows the schematic of a straight waveguide using the L-interface boundary. We define the radius of the large air holes at the boundary as $r_e$. By tuning the $r_e$, the working band of each waveguide can be controlled. As shown in Fig. 3(b), the decrease of $r_e$ (120 nm, 116 nm, 110 nm) results in a redshift of the working bands due to the increased effective refractive index of the edge states.

We performed numerical simulations on 11 sets of straight waveguides using RCP incident light, with $r_e$ gradually decreasing from 120 nm to 110 nm at a step of 1 nm. The transmittance spectra are shown in Fig. 3(c). The tuning aims to separate the working bands of INA and INB to minimize interference and achieve a stable output. According to the definition of logic "1", the working band is defined as where the forward transmittance is higher than 0.7 (TF > 0.7). The plot of the working band versus different radii is shown in Fig. 3(d). The working band is quite sensitive to the radius change, which presents a linear relationship with $r_e$ (the slope is 5.792). Therefore, when $r_e$ decreases by 1 nm, the central wavelength of the working band increases by 5.792 nm. We find three $r_e$ values meeting our requirements, which are $r_e$ = 120 nm (the working band is 1476 nm - 1512 nm), $r_e$ = 116 nm (the working band is 1494.7 nm - 1542 nm), and $r_e$ = 110 nm (the working band is 1522 nm - 1582 nm), respectively. As one can see, the working bands in the cases of $r_e$ = 120 nm and $r_e$

= 110 nm can be well separated. Meanwhile, the working band in the case of $r_e$ = 116 nm overlaps with both $r_e$ = 120 nm and $r_e$ = 110 nm. Therefore, we use $r_1$ = 120 nm for INA and $r_2$ = 110 nm for INB, respectively. Meanwhile, $r_3$ = 116 nm is used for OUT1. As a result, inputs from INA and INB can highly efficiently transmit through OUT1. In addition, to minimize the light leakage from OUT2, the air holes at the boundary are removed. Thus, the transmittance of OUT2 approaches zero in the range of 1450 nm - 1530 nm. (The detail is shown in Supplementary material S2).

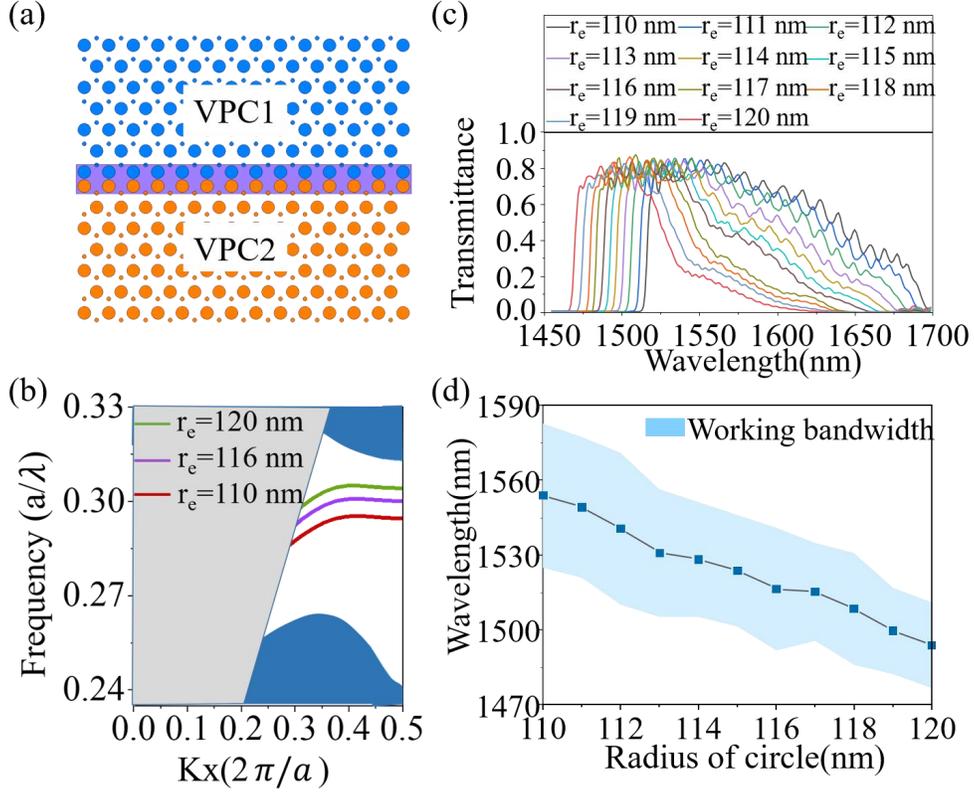

**Fig. 3.** (a) A schematic diagram of a straight waveguide based on the L-interface boundary (the big circles in the purple area are the air holes at the boundary, the radius is $r_e$). (b) Band diagram of the valley-dependent edge states (the green, purple and red lines correspond to $r_e$ = 120 nm, $r_e$ = 116 nm, and $r_e$ = 110 nm, respectively). (c) Transmittance spectra of straight waveguides with different $r_e$. (d) The plot of working band versus different $r_e$.

We design Z-shape waveguides with INA or INB combined with OUT1 to study the transmission properties of different combinations. The schematics of the Z-shape waveguides are shown in Figs. 4(a) and (b), respectively. The green and red circles in the Z1 and Z2 waveguides (Figs. 4(a) and (b)) have $r_1$ = 120 nm and $r_2$ = 110 nm, respectively. The purple circles in both waveguides have an $r_3$ = 116 nm. The transmittance spectra are plotted in Fig. 4(e). Here we choose two working wavelengths for INA and INB to achieve the function of an AOLG, which are $\lambda_1$ = 1490 nm and $\lambda_2$ = 1528 nm (marked by the dashed lines in Fig. 4(e)), respectively. Those wavelengths are chosen due to the high transmittance (TF > 0.75). The intensity distributions in the Z-shape waveguides at the two working wavelengths in Figs. 4(c) and (d) show that the light waves are well confined within the waveguides.

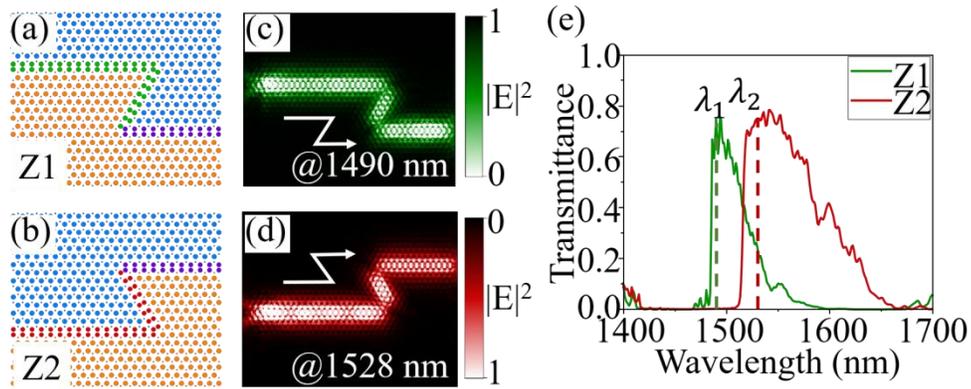

**Fig. 4.** Z-shape waveguide designs and transmittance. (a) Schematic of Z1 waveguide composed of INA and OUT1. (b) Schematic of Z2 waveguide composed of INB and OUT1. (c) The electric field intensity distributions in Z1 waveguide at $\lambda_1 = 1490$ nm. (d) The electric field intensity distributions in Z2 waveguide at $\lambda_2 = 1528$ nm; (e) Transmittance spectra of the two Z-shape waveguides.

## 4. Performance analysis of the designed all-optical OR gate

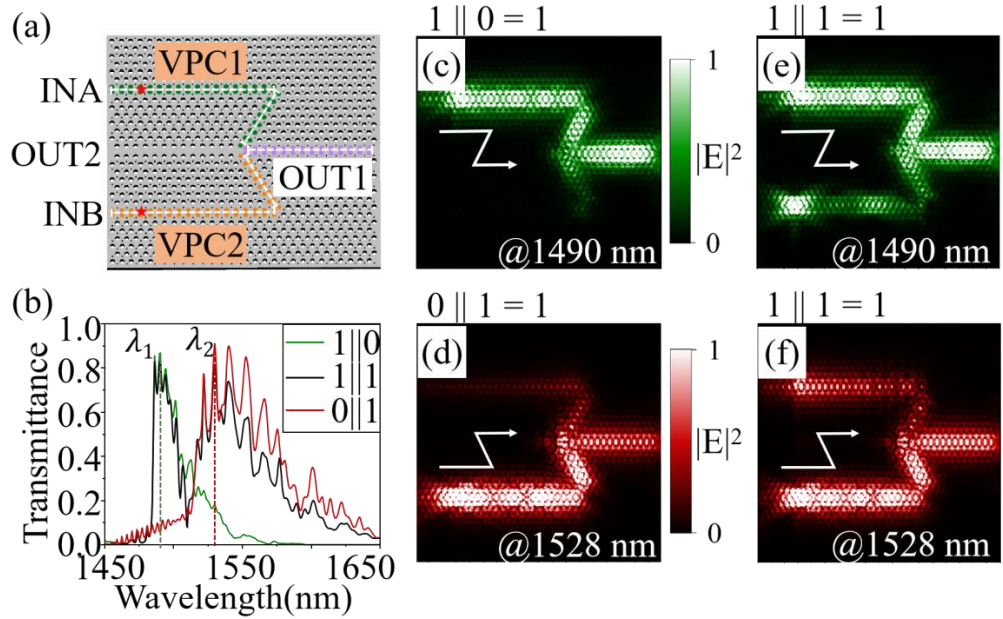

**Fig. 5.** The designed AOLG and its function. (a) 3D structural diagram of the ALOG gate. (b) Transmittance spectra of the AOLG in three different cases. Electric field intensity distributions at the wavelength of $\lambda_1 = 1490$ nm input from INA alone (c) (1 || 0 = 1) and both INA and INB (e) (1 || 1 = 1), respectively. Electric field intensity distributions at the wavelength of $\lambda_2 = 1528$ nm input from INB alone (d) (0 || 1 = 1) and both INA and INB (f) (1 || 1 = 1), respectively.

We then combine the two Z-shape waveguides to form the all-optical OR gate, shown in Fig. 5(a). The transmittance spectra of the AOLG are shown in Fig. 5(b), in which one can see different input cases from INA (1 || 0), INB (0 || 1), and INA+INB (1 || 1). In all three cases, the output from OUT1 has a transmittance higher than 0.7 at either $\lambda_1 = 1490$ nm or $\lambda_2 = 1528$ nm. The corresponding intensity distributions are plotted in Figs. 5(c)-(f). When the light waves at the designed working wavelengths are input from INA (Fig. 5(c)) and INB (Fig.

5(d)) separately, they can highly efficiently transmit through the AOLG. The transmittances are 0.86 and 0.90 at the wavelengths of $\lambda_1 = 1490$ nm or $\lambda_2 = 1528$ nm, respectively. In comparison, due to the well-separated work bands of INA and INB, when light waves at the two wavelengths are input from both INA and INB (1 || 1) (Figs. 5(e) and (f)), they don't interfere. Because the light waves at $\lambda_1 = 1490$ nm can not transmit through INB. Similarly, the light waves at $\lambda_2 = 1528$ nm have very low transmittance in INA. Here we notice that there is a small portion of light transmitting from one input to the other input (for example, from INA to INB at $\lambda_1 = 1490$ nm and from INB to INA at $\lambda_2 = 1528$ nm) at the junction of INA, INB, and OUT1 (shown in Figs. 5(c) and (d)). However, this effect doesn't influence the overall performance of the AOLG, as shown in Table 1.

Table 1. All-optical logic or gate truth table

| INA | INB | Logic output | Transmittance |
|---|---|---|---|
| 0 | 0 | 0 | 0 |
| 1 | 0 | 1 | 0.86 |
| 0 | 1 | 1 | 0.90 |
| 1 | 1 | 1 | 0.82 |

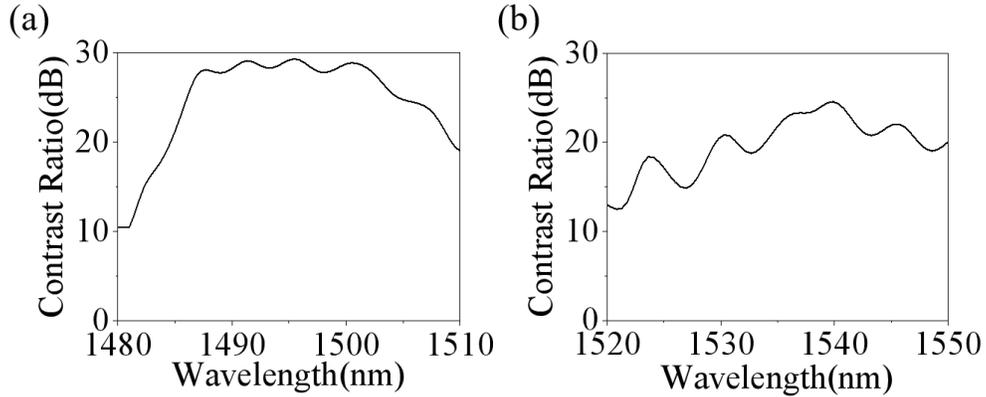

Fig. 6. Contrast ratio plots when light is only input from INA (a) and INB (b).

The performance of the AOLG can be further studied by using the logic gate contrast function, which is defined as $R = 10\log\left(\frac{P_1}{P_0}\right)$, where $P_1$ and $P_0$ are the transmittances of the output and the unexcited input, respectively. Most applications desire a high contrast ratio, which is challenging based on reported designs. The contrast ratio curves are shown in Figs. 6(a) and (b) for the inputs from INA and INB, respectively. The wavelength ranges are decided according to the working bands of INA and INB. The contrast ratios are 28.8 dB at $\lambda_1 = 1490$ nm in INA and 20.5 dB at $\lambda_2 = 1528$ nm in INB, which are much higher compared to state-of-the-art AOLGs [27,29,30,33].

To show the advantages of using two well-separated working bands for the two inputs, we also perform the simulation on the logic gate composed of identical straight waveguides (details in Supplementary material S3). In this case, the transmittance in each case is relatively lower compared to the separate design. That is due to the light transmitting from one input (INA or INB) to the other (INB or INA). Because both INA and INB can support the same edge states. The situation can be avoided by using different working bands for INA and INB, as demonstrated in this work. In this way, the forward transmittance of each input can be maximized, which is essential for any AOLGs for quantum computing since single photon transmission is necessary. Further, to demonstrate our design can work with different spin states (LCP or RCP light) due to the spin-valley locking effect, we perform the simulation on the input with different handedness (details in Supplementary material S4). Here LCP and RCP light is defined as logic "1" and "0", respectively. The output transmittance (TF > 0.7) is still designated as logic "1". Due to the spin-valley locking effect, the LCP and RCP light propagates in opposite directions, resulting in high and low transmittances corresponding to the logic "1" and "0" states, respectively. In this way, the chirality of the logic gate allows using different spin states as different logic states, which is achieved among various AOLGs based on PC structures for the first time to the best of our knowledge.

## 5. Conclusion

We have demonstrated a high-performance AOLG design based on a VPC structure. By tuning the working bands of the two inputs, we minimize the interference and cross-coupling between the inputs to ensure high transmittance output (up to 0.9) and high contrast ratio (up to 28.8 dB). Moreover, the chiral response of the logic gate due to the spin-valley locking effect allows using different spin states as inputs for quantum computing. The device's footprint is as small as $18 \times 12$ μm$^2$, which is suitable for high-density on-chip integration. In addition, the device is compatible with the CMOS nanofabrication technique. The design principle can be generally applied to design other photonic devices based on VPCs for broad applications.

### Funding

This work was supported by the National Natural Science Foundation of China (Grant No. 11904255), the Key R&D Program of Shanxi Province (International Cooperation) (201903D421052).

### Declaration of Competing Interest

The authors declare that they have no known competing financial interests or personal relationships that could have appeared to influence the work reported in this paper.

### Data availability

No data was used for the research described in the article.

### Appendix A. Supplementary material

See Supplement material for supporting content.

### References

1. N. C. Panoiu, M. Bahl, and R. M. Osgood, "All-optical tunability of a nonlinear photonic crystal channel drop filter," Opt Express, **12**(8), 1605-1610,(2004).
2. V. Gomez et al. "Concatenated logic functions using nanofluidic diodes with all-electrical inputs and outputs," Electrochem Commun, **88**, 52-56, (2018).

3. F. Chi, Q. Xu, and X. Long, "Editorial: physical model and applications of high-efficiency electro-optical conversion devices," Front Phys, **9**, 819303, (2021).
4. W. Fu, M. R. Xu, X. W. Liu, C. L. Zou, C. C. Zhong, X. Han, M. H. Shen, Y. T. Xu, R. S. Cheng, S. H. Wang, L. Jiang, and H. Tang, "Cavity electro-optic circuit for microwave-to-optical conversion in the quantum ground state," Phys Rev A, **103**(5), 053504, (2021).
5. D. M. Kong, Y. Li, H. Wang, X. P. Zhang, J. Y. Zhang, J. Wu, and J. T. Lin., "All-optical XOR gates for QPSK signals based on four-wave mixing in a semiconductor optical amplifier," IEEE Photonic Tech L, **24**(12), 988-990, (2012).
6. S. Z. Ma, Z. Chen, H. Z. Sun, and N. K. Dutta., "High speed all optical logic gates based on quantum dot semiconductor optical amplifiers," Opt Express, **18**(7), 6417-6422, (2010).
7. A. Kotb, K. E. Zoiros, and W. Li, "Realization of ultrafast all-optical NAND and XNOR logic functions using carrier reservoir semiconductor optical amplifiers," J Supercomput, **77**(12), 14617-14629, (2021).
8. J. Wang, J. Q. Sun, X. L. Zhang, D. X. Huang, and M. M. Fejer., "All-Optical format conversions using periodically poled lithium niobate waveguides," IEEE J Quantum Elect, **45**(2), 195-205, (2009).
9. J. Wang, J. Q. Sun, Q. Z. Sun, D. L. Wang, X. L. Zhang, D. X. Huang, and M. M. Fejer, "PPLN-based flexible optical logic and gate," IEEE Photonic Tech L, **20**(3), 211-213, (2008).
10. Y. Y. Xie, Y. Y. Yin, T. T. Song, Y. C. Zhu, J. X. Chai, B. C. Liu, and Y. C. Ye, "The design and simulation of a multifunctional logic device based on plasmon-induced transparency using two semicircular resonators," Optik, **255**, (2022).
11. F. K. Law, M. R. Uddin, A. T. C. Chen, and B. Nakarmi, "Positive edge-triggered JK flip-flop using silicon-based micro-ring resonator," Opt Quant Electron, **52**(6), 314, (2020).
12. L. Zhang, J. F. Ding, Y. H. Tian, R. Q. Ji, L. Yang, H. T. Chen, P. Zhou, Y. Y. Lu, W. W. Zhu, and R. Min, "Electro-optic directed logic circuit based on microring resonators for XOR/XNOR operations," Opt Express, **20**(11), 11605-11614, (2012).
13. L. P. Caballero, M. L. Povinelli, J. C. Ramirez, P. S. S. Guimaraes, and O. P. V. Neto, "Photonic crystal integrated logic gates and circuits," Opt Express, **30**(2), 1976-1993, (2022).
14. A. Sharma, K. Goswami, H. Mondal, T. Datta, and M. Sen, "A review on photonic crystal based all-optical logic decoder: linear and nonlinear perspectives," Optical and Quantum Electronics, **54**(2), (2022).
15. H. M. E. Hussein, T. A. Ali, and N. H. Rafat, "A review on the techniques for building all-optical photonic crystal logic gates," Opt Laser Technol, **106**, 385-397, (2018).
16. H. M. Fei, Q. Zhang, M. Wu, H. Lin, X. Liu, Y. B. Yang, M. D. Zhang, R. Guo, and X. T. Han, "Asymmetric transmission of light waves in a photonic crystal waveguide heterostructure with complete bandgaps," Appl Opt, **59**(14), 4416-4421, (2020).
17. H. M. Fei, S. Yan, M. Wu, H. Lin, Y. B. Yang, M. D. Zhang, and X. T. Han, "Photonic crystal with 2-fold rotational symmetry for highly efficient asymmetric transmission," Opt Commun, **477**, 126346, (2020).
18. H. M. Fei, M. Wu, T. Xu, H. Lin, Y. B. Yang, X. Liu, M. D. Zhang, and B. Z. Cao, "A broadband polarization-insensitive on-chip reciprocal asymmetric transmission device based on generalized total reflection principle," J. Opt. **20**(9), 095004 (2018).
19. H. M. Fei, M. Wu, H. Lin, X. Liu, Y. B. Yang, M. D. Zhang, and B. Z. Cao, "An on-chip nanophotonic reciprocal optical diode for asymmetric transmission of the circularly polarized light," Superlatt. Microstruct. **132**, 106155 (2019).
20. H. M. Fei, M. Wu, H. Lin, Y. B. Yang, X. Liu, M. D. Zhang, and B. Z. Cao, "A highly efficient asymmetric transmission device for arbitrary linearly polarized light," Photonics. Nanostruct, **41**, 100829, (2020).
21. M. Wu, H. M. Fei, H. Lin, X. D. Zhao, Y. B. Yang, and Z. H. Chen, "Design of asymmetric transmission of photonic crystal heterostructure based on two-dimensional hexagonal boron nitride material," Acta Phys. Sin., **70**(2), 028501, (2021).
22. W. Q. Zhi, H. M. Fei, Y. H. Han, M. Wu, M. D. Zhang, X. Liu, B. Z. Cao, and Y. B. Yang, "Unidirectional transmission of funnel-shaped waveguide with complete bandgap," Acta Phys. Sin., **71**(3), 038501, (2022).
23. H. M. Fei, T. Xu, X. Liu, H. Lin, Z. H. Chen, Y. B. A. Yang, M. D. Zhang, B. Z. Cao, and J. Q. Liang, "Interface of photonic crystal heterostructure for broadening bandwidth of unidirectional light transmission," Acta Phys. Sin., **66**(20), 204103, (2017).
24. H. M. Fei, J. J. Wu, Y. B. Yang, X. Liu, and Z. H. Chen, "Magneto-optical isolators with flat-top responses based on one-dimensional magneto-photonic crystals," Photonic. Nanostruct., **17**, 15-21, (2015).
25. H. M. Fei, S. Yan, Y. C. Xu, H. Lin, M. Wu, Y. B. Yang, Z. H. Chen, Y. Tian, and Y. M. Zhang, "Photonic crystal heterostructure with self-collimation effect for broad-band asymmetric optical transmission," Acta Phys. Sin., **69**, 184214, (2020).
26. R. R. Fan, X. L. Yang, X. F. Meng, and X. W. Sun, "2D photonic crystal logic gates based on self-collimated effect," J Phys D Appl Phys, **49**(32), 325104, (2016).
27. M. R. Jalali-Azizpoor, M. Soroosh, and Y. Seifi-Kavian, "Application of self-collimated beams in realizing all-optical photonic crystal-based half-adder," Photonic Netw Commun, **36**(3), 344-349, (2018).
28. W. Liu, D. Yang, G. Shen, H. Tian, and Y. F. Ji, "Design of ultra compact all-optical XOR, XNOR, NAND and OR gates using photonic crystal multi-mode interference waveguides," Opt Laser Technol, **50**, 55-64, (2013).
29. C. R. Tang, X. Dou, Y. Lin, H. Yin, B. Wu, and Q. Zhao, "Design of all-optical logic gates avoiding external phase shifters in a two-dimensional photonic crystal based on multi-mode interference for BPSK signals," Opt Commun, **316**, 49-55, (2014).

30. H. M. E. Hussein, T. A. Ali, and N. H. Rafat, "New designs of a complete set of Photonic Crystals logic gates," Opt Commun, **411**, 175-181, (2018).
31. D. G. S. Rao, S. Swarnakar, V. Palacharla, K. S. R. Raju, and S. Kumar, "Design of all-optical AND, OR, and XOR logic gates using photonic crystals for switching applications," Photon Netw Commun, **41**(1), 109–118, (2021).
32. H. Alipour-Banaei, S. Serajmohammadi, and F. Mehdizadeh, "All optical NAND gate based on nonlinear photonic crystal ring resonators," Optik, **130**, 1214-1221, (2017).
33. M. Ghadrdan and M. A. Mansouri-Birjandi, "Concurrent implementation of all-optical half-adder and AND & XOR logic gates based on nonlinear photonic crystal," Opt Quant Electron, **45**(10), 1027-1036, (2013).
34. Y. T. Yang, Y. F. Xu, T. Xu, H. X. Wang, J. H. Jiang, X. Hu, and Z. H. Hang, "Visualization of a Unidirectional Electromagnetic Waveguide Using Topological Photonic Crystals Made of Dielectric Materials," Phys Rev Lett, **120**(21), 217401, (2018).
35. S. A. Skirlo, L. Lu, Y. C. Igarashi, Q. H. Yan, J. Joannopoulos, and M. Soljacic, "Experimental observation of large chern numbers in photonic crystals," Phys Rev Lett, **115**(25), 253901, (2015).
36. F. Gao, H. R. Xue, Z. J. Yang, K. F. Lai, Y. Yu, X. Lin, Y. D. Chong, G. Shvets, and B. L. Zhang, "Topologically protected refraction of robust kink states in valley photonic crystals," Nat Phys, **14**(2), 140-144, (2018).
37. J. C. Lu, X. D. Chen, W. M. Deng, M. Chen, and J. W. Dong, "One-way propagation of bulk states and robust edge states in photonic crystals with broken inversion and time-reversal symmetries," J Opt, **20**(7), 075103, (2018.).
38. M. C. Rechtsman, J. M. Zeuner, Y. Plotnik, Y. Lumer, D. Podolsky, F. Dreisow, S. Nolte, M. Segev, and A. Szameit, "Photonic floquet topological insulators," Nature, **496**(7444), 196-200, (2013).
39. Y. H. Han, H. M. Fei, H. Lin, Y. M. Zhang, M. D. Zhang, and Y. B. Yang, "Design of broadband all-dielectric valley photonic crystals at telecommunication wavelength," Opt. Commun. **488,** 126847 (2021).
40. M. Wu, Y. B. Yang, H. M. Fei, H. Lin, Y. H. Han, X. D. Zhao, and Z. H. Chen, "Unidirectional transmission of visible region topological edge states in hexagonal boron nitride valley photonic crystals," Opt Express, **30**(4), 6275-6283, (2022).
41. X. D. Chen, F. L. Shi, H. Liu, J. C. Lu, W. M. Deng, J. Y. Dai, Q. Cheng, and J. W. Dong, "Tunable Electromagnetic Flow Control in Valley Photonic Crystal Waveguides," Physical Review Applied, **10**(4), 044002, (2018).
42. R. Prevedel, P. Walther, F. Tiefenbacher, P. Bohl, R. Kaltenbaek, T. Jennewein, and A. Zeilinger, "High-speed linear optics quantum computing using active feed-forward," Nature, **445**(7123), 65-9, (2007).